\documentclass[12pt]{article}
\usepackage{amsthm}
\usepackage{amsmath,amscd}
\usepackage{amsfonts}
\usepackage[dvips]{graphicx}

\newtheorem{proposition}{Proposition}[section]

\newtheorem{acknowledgment*}{Acknowledgment}

\numberwithin{equation}{section}

\newcommand{\be}{\begin{equation}}
\newcommand{\ee}{\end{equation}}
\newcommand{\bd}{\begin{displaymath}}
\newcommand{\ed}{\end{displaymath}}

\newcommand{\R}{\mathbb R}

\renewcommand{\vec}[1]{\boldsymbol{#1}}

\begin{document}
\Large
\begin{center}\bf{ Diffusion's induced transport in periodic channels and  an  inverse problem}\end{center}
  \normalsize
\small \begin{center} Gershon Wolansky\footnote{
Department of Mathematics,  Technion, Haifa, Israel. E.mail gershonw@math.technion.ac.il}\end{center}
\normalsize
\begin{abstract} A diffusion's induced transport is defined for a linear model of a Fokker-Plank equation under periodic boundary conditions in one-dimensional geometry.  The flow is generated by a diffusion and a periodic deriving force induced by a velocity potential. An inverse problem is suggested for evaluating the deriving force in terms of the response function associated with the flow. It is also shown that the inverse problem can be partially solved under some simplifying assumption.
\end{abstract}
\section{Transport induced by a diffusion}
Recently, there is an increasing interest in
transport process  induced by an irreversible  diffusion.
This is motivated by some  aspects of molecular dynamics and molecular motors [DEO].

The general paradigm of diffusion's induced  transport is a model of  several components. Each component undergo an independent diffusion process,
and the  different components are  coupled together by the poisson process  [SKB], [PJAP].

In some cases  the basic model is that of a diffusion of a single component, while the transport mechanism is induced by an time periodic deriving potential [KK]:
\be\label{model} \rho_t=\left( \sigma\rho_x + \Psi_x\rho\right)_x \ \ \ \ x\in [0,1] \ , \   \ee
\be\label{model1}(\sigma_x + \Psi_x\rho)_{(0,t)}= (\sigma_x + \Psi_x\rho)_{(1,t)}=0 \ . \ee
 Here the deriving potential $\Psi:=\Psi(x,t)=\Psi(x,t+T)$ is an appropriately defined function, and $\rho\geq 0$ is the probability density of position of a "test particle".

The model system (\ref{model}), as the one studied in [KK],  deals  with no-flux boundary conditions (\ref{model1}).  It is proved that the solution of (\ref{model},\ref{model1}) converges asymptotically, under general condition, to a periodic solution $\overline{\rho}(x,t)=\overline{\rho}(x,t+T)$. The phenomenon of transport is introduced in a descriptive way: The period average of the asymptotic density $\overline{\rho}$ is shown to concentrate, under suitable conditions on $\Psi$, near one of the ends of the interval.

The basic model introduced in  [DEO], however, implies  a geometry of periodic spatial structure. So, it seems reasonable to replace  (\ref{model1})
by
\be\label{model2} \rho(0,t)=\rho(1,t) \ \ \ , \ \ \ \rho_x(0,t)=\rho_x(1,t) \ . \ee
where $\Psi$ is assumed to be $1-$periodic in space (in addition to its time periodicity).
Under these b.c., there is a natural definition for a diffusion transport, as follows:

Consider a solution $\rho=\rho(x,t)$ of (\ref{model}, \ref{model2}). The velocity of a "particle" driven by this process
is composed of the driving potential $-\Psi_x$ and the {\it osmotic velocity} $-\sigma\rho_x/\rho$:
 \be\label{vdef} \vec{v}(x,t)=-\Psi_x-\sigma\rho_x/\rho\ee
 so the diffusion equation (\ref{model}) takes the form of the {\it continuity equation}
 $$ \frac{\partial\rho}{\partial t} + \frac{\partial(\vec{v}\rho)}{\partial x}=0 \ . $$
 Hence, we may describe the orbit $x=x(t)$ of such a particle by
\be\label{vel} \frac{d x}{dt} = \vec{v}(x(t),t) \ . \ee
Recall that $\vec{v}$ is $1-$periodic in $x$ and $T-$periodic in $t$.
 The {\it mean velocity}, given by
$$ \kappa:= \lim_{t\rightarrow\infty}\frac{x(t)}{t}= \lim_{t\rightarrow\infty} \frac{1}{t}\int^t\vec{v}(x(s),s)ds$$ is independent of its initial position $x(0)$ and the initial data $\rho(, 0)$. It is just natural to define this process as a {\it transport process} if $\kappa\not= 0$.

How can we prove that a certain driving potential $\Psi$ induces a transport $\kappa\not= 0$? As a case study we may consider a spatial periodic function
$\psi(x)=\psi(x+1)$ and  \begin{equation}\label{cs}\Psi(x,t):=\psi(x-V t)\ee for some $V\in\R$ ({\it applied voltage}). This way we get a time periodic deriving  potential of temporal period
$T=1/V$ and spatial period $1$.

Switching now to the variable $z=x-V t$, (\ref{model}, \ref{model2}) takes the form
\be\label{mode2} \rho_t=\left( \sigma\rho_z + \psi_z\rho+V\rho\right)_z \ , \  \ \ \rho(z,t)-\rho(z+1,t)= \rho_z(z,t)-\rho_z(z+1,t)=0 \ee
where $\rho(z,t)$ stands for $\rho(x-V t,t)$. In this case, the asymptotic limit $\overline{\rho}$ is just the stationary solution
of (\ref{mode2}) which implies that the {\it current} $I$ is a constant:
\be\label{cf} \sigma \overline{\rho}_{z}+ \psi_z\overline{\rho}+ V\overline{\rho}=I \ . \ee
This current  is uniquely determined from $V$ and the normalization condition
\be\label{nor} \int_0^1\overline{\rho}=\int_0^1\rho_0=1  \ . \ee
From (\ref{vel}, \ref{vdef}) and (\ref{cf})
$$ \frac{dz}{dt}= -\frac{I}{\overline{\rho}(z)}$$
so from (\ref{nor})
$$\lim_{t\rightarrow\infty} \frac{z(t)}{t}=- I$$
and
$$ \kappa=V-I \ .  $$
Now, divide (\ref{cf}) by $\overline{\rho}$ and integrate over one period. Since both $\psi$ and $\ln(\overline{\rho})$ are periodic, we obtain
 $$ \kappa=I\int_0^1(\overline{\rho})^{-1}$$
 hence
 $$ \kappa=V\left( 1-\frac{1}{\int_0^1(\overline{\rho})^{-1}} \right) \ . $$    
 We now recall from (\ref{nor}) that $\int_0^1(\overline{\rho})^{-1} \geq 1$, and $\int_0^1(\overline{\rho})^{-1} = 1$ if and only if $\overline{\rho}\equiv 1$, which is the case if and only if $\psi$ is a constant. We obtained
 \begin{proposition}
 The process (\ref{model}, \ref{model2}) under condition (\ref{cs}) is a transport process (namely $\kappa\not= 0$) if and only if $V\not= 0$ and $\psi$ is not a constant. Moreover, the mean velocity $\kappa$ is always in the direction of the applied voltage $V$, and $|\kappa|\leq |V|$.
 \end{proposition}
\section{The inverse Problem}
In most applications we cannot measure directly the details of the driving potential $\psi$. What we can measure is the current $I$ as a function of the applied voltage $V$ and the "temperature" $\sigma$.

The question we address now is related to the model problem (\ref{mode2}):
\vskip .2in\noindent
{\it What can be said about the driving potential $\psi$ from a complete knowledge of the mean velocity $\kappa(V,\sigma)$ (equivalently, the {\it response current} $I(V, \sigma)$), for each applied voltage $V\in\R$ and positive temperature $\sigma>0$ ?}

Indeed, it is shown in the appendix below that
\be\label{form1}{\cal L}_{z}^{-1}\left(\frac{e^V-1}{I(\sigma V,\sigma)}\right)= \left\{ \begin{array}{c}
                                                                              \int_0^1e^{ \sigma^{-1}(\psi(x)- \psi(x+z))}dx \ \ \ \text{for} \ \ z\in[-1,0] \\
                                                                              0 \ \ \text{otherwise}
                                                                            \end{array}\right.
 \ee
 where  ${\cal L}^{-1}_z$ is the inverse Laplace transform action on functions of $V$ into functions of $z$.

So, a complete knowledge of the   response current $I$ as a function of $\sigma$ and $V$ yields a complete knowledge of the function
    $$ F(\sigma,z):=    \int_0^1e^{ \sigma^{-1}(\psi(x)- \psi(x+z))}dx   $$
for any $\sigma>0$ and $z\in [0,1]$. I conjecture that a complete knowledge of $F$ yields a complete knowledge on $\psi$, up to and additive constant and shift. 
  
  At this stage, however, I can only prove that a {\it partial knowledge} of  $I(V,\sigma)$ yields, under certain assumption,  a partial information on $\psi$:
   
  At the limit $V=0$, the {\it resistance} $ (\partial I/\partial V)^{-1}(0,\sigma)$ is given by
\be\label{fr} \int_{-1}^0\int_0^1e^{ \sigma^{-1}(\psi(x)- \psi(x+z))}dxdz = \lim_{V\rightarrow 0} \left(\frac{e^V-1}{I(\sigma V,\sigma)}\right)
=\sigma^{-1}\left(\frac{\partial I}{\partial V}\right)^{-1}_{(0,\sigma)}\ee
\begin{proposition}
Assume $\psi$ is anti-symmetric $\psi(x)=-\psi(-x)$ (in addition to being $1-$periodic). Assume further we can measure the resistance $(dI/dV)^{-1}$ as a function of the temperature $\sigma$ at zero voltage $V=0$.  Then we can find the distribution of $\psi$.
\end{proposition}
Indeed, if we know $(\partial I/\partial V)_{V=0}^{-1}$ as a function of  $\sigma>0$, then we can expand this function as a power series at $\sigma=\infty$ to obtain
\be\label{ch1} \sigma^{-1}(dI/dV)^{-1} =   1 + \frac{c_1}{\sigma} + \frac{c_2}{\sigma^2} + \ldots  \ . \ee
Then, we expand the exponent on the left of (\ref{fr}) in powers of $\sigma^{-1}$:
\be\label{ch2} \int_{-1}^0\int_0^1e^{\sigma^{-1}(\psi(x)- \psi(x+z))}dxdz =1+ \sum_{k=1}^\infty \frac{\sigma^{-k}}{k!}\int_{-1}^1\int_0^1(\psi(x)- \psi(x+z))^kdxdz \ee
Recall that $\psi$ is $1-$periodic. Comparing equal powers of $\sigma$ between (\ref{ch1}) and (\ref{ch2}) we get
$c_1=0$ and, for $k>1$,
$$ c_k=(k!)^{-1}\sum_{j=0}^k  (-1)^j\left(
                             \begin{array}{c}
                               k \\
                               j \\
                             \end{array}
                           \right)M_j(\psi)M_{k-j}(\psi)
$$
where $M_j(\psi)=\int_0^1\psi^j$, the $jth$ moment of $\psi$.  By anti-symmetry of $\psi$ we get $M_j(\psi)=0$ for odd $j$. This implies, in particular, that $c_k=0$ for odd $k$. In addition, the assumed known values of $c_k$ for even $k$ determine all even moments of $\psi$, hence its distribution.

\vskip .2in\noindent
{\bf Appendix}

 We first calculate $I=I(V,\sigma)$ from (\ref{cf}). Without limitation to generality we assume $\int_0^1\psi=0$.
 Let
 $$ f_+(z):=\exp\left(\frac{V z + \psi(z)}{\sigma}\right) \ \ ; \ \ f_-(z):=\exp\left(-\frac{V z + \psi(z)}{\sigma}\right)$$
 and
 $$ F_\pm(z):=\int_0^z f_\pm(s)ds \ . $$
 The solution of (\ref{cf}) takes the form
 $$ \overline{\rho}(z)=f_-(z)\left[ \beta +I F_+(z)\right]$$
 where $I$ and $\beta$ are determined from the normalization (\ref{nor}) and periodicity condition of $\overline{\rho}$ as follows:
 \begin{equation}\label{1} \overline{\rho}(0)=\overline{\rho}(1) \ \ \ \Longrightarrow \beta(1- e^{-V/\sigma}) -I e^{-V/\sigma}F_+(1) =0\end{equation}
 \begin{equation}\label{2} \int_0^1\overline{\rho}=1 \ \ \ \Longrightarrow \beta F_-(1)+I\int_0^1 f_-F_+ =1\end{equation}
 From (\ref{1}, \ref{2}) we factor out $I$ to obtain
 $$ I=\frac{1-e^{-V/\sigma}}{e^{-V/\sigma} F_+(1)F_-(1)+ (1-e^{-V/\sigma})\int_0^1f_-F_+}$$
  Integration by parts yields
 $$F_+(1) F_-(1) = \int_0^1 f_-F_+ + \int_0^1 f_+F_- \ . $$
 Let $\int_0^1 F_+f_- = H_+(V, \sigma)$, $\int_0^1 F_-f_+ = H_-(V, \sigma)$. Then
 \be\label{t} I(V,\sigma)=\frac{1-e^{-V/\sigma}}{e^{-V/\sigma} H_+ (V,\sigma)+ H_-(V, \sigma)} \ . \ee
 Conversely,
 \begin{equation}\label{3}e^{-V/\sigma} H_-(V,\sigma)+ H_+(V,\sigma)=\frac{1-e^{-V/\sigma}}{I(V,\sigma)} \ . \end{equation}
Let now
 $$h_+(z):= \left\{ \begin{array}{cc}
              \int_z^1 e^{ \sigma^{-1}(\psi(x-z)- \psi(x))}dx & z\in [0,1] \\
              0 & z\not\in [0,1]
            \end{array} \right.
 $$
 $$h_-(z):= \left\{ \begin{array}{cc}
              \int_{-z}^1 e^{ \sigma^{-1}(\psi(x)- \psi(x+z))}dx & z\in [-1,0] \\
              0 & z\not\in [0,1]
            \end{array} \right.
 $$
 In particular (taking into account the periodicity of $\psi$)
 \be\label{h+-} h_+(z+1) + h_-(z)={\bf 1}_{[-1,0]}(z)\int_0^1e^{ \sigma^{-1}(\psi(x)- \psi(x+z))}dx\ee
where ${\bf 1}_{[-1,0]}$ is the indicator function of $[-1,0]$.
 We observe that
 $$H_+(V)=\int_{-\infty}^\infty e^{-V z/\sigma} h_+(z) dz \ \ \ \ , \ \ \ H_-(V)=\int_{-\infty}^\infty e^{-V z/\sigma} h_-(z) dz  . $$
 By (\ref{t}, \ref{3}) we get
 $$ \int_{-\infty}^\infty e^{-Vz/\sigma}(h_-(z)+h_+(z+1)) dz = \frac{e^{V/\sigma}-1}{I(V, \sigma)} \ . $$
So, if we take the inverse Laplace transform of (\ref{3}) (as function of $V/\sigma$) we obtain from (\ref{h+-})
$$ \int_0^1e^{ \sigma^{-1}(\psi(x)- \psi(x+z))}dx= {\cal L}_{z}^{-1}\left(\frac{e^V-1}{I(\sigma V,\sigma)}\right)   $$
provided $z\in [-1,0]$, and ${\cal L}_{z}^{-1}\left(\frac{e^V-1}{I(\sigma V,\sigma)}\right)=0$ otherwise.
\begin{center}{\bf References}\end{center}
\begin{description}
\item{[DEO]} C. Doering, B. Ermentrout and G. Oster, {\it Rotary DNA motors}, Biophysical Journal, 69, (1995), 2256--2267
\item{[PJAP]} A. Parmeggiani, F. Ju¨licher, A. Ajdari and J. ProstEnergy: {\it  transduction of isothermal ratchets: Generic aspects and specific examples
close to and far from equilibrium}, Phys. Rev E, 60, \#2, (1999), 2127--2140
\item{[SKB]} S.  Hastings, J.B.  McLeod,
 D. Kinderlehrer: \ {\it Diffusion mediated transport in multiple state systems},  SIAM J. Math. Anal. 39 (2007/08), no. 4, 1208--1230
\item{[KK]} D. Kinderlehrer and M. Kowalczyk: {\it Diffusion-Mediated Transport
and the Flashing Ratchet}, Arch. Rational Mech. Anal. 161 (2002) 149–179
\end{description}
\end{document}